\documentclass[shortnote,onecolumn]{jpsj3} 
\usepackage{color}

\title{Nonequilibrium Extension of Onsager Relations for Thermoelectric Effects of Mesoscopic Conductors}

\author{
Eiki \textsc{Iyoda}$^{1}$
\thanks{E-mail address: iyoda@issp.u-tokyo.ac.jp},
Yasuhiro \textsc{Utsumi}$^{1,2}$,
and Takeo \textsc{Kato}$^{1}$
}

\inst{$^{1}$ Institute for Solid State Physics, the University of Tokyo, Chiba 277-8581, Japan \\
$^{2}$ CREST, Japan Science and Technology (JST), Saitama, 332-0012, Japan}

\kword{
fluctuation theorem, 
Onsager relation, 
nonequilibrium quantum transport
}

\begin{document}
\maketitle

It is well known that the Onsager reciprocal relations give important identities
between transport coefficients for heat and charge conduction
in the linear response regime.~\cite{Onsager31,Callen85} For example, the Peltier coefficient $\Pi$ is 
related to the Seebeck coefficient $S$ in a simple form as $S = \Pi/T$, where $T$ is the temperature.
The Onsager-Casimir symmetry, which is the extension of the Onsager relation in the presence 
of an external magnetic field, is also derived from the microscopic reversibility.~\cite{Casimir45} 
Recently, a general relation called the steady-state fluctuation theorem has been derived
by the same concept of the microscopic reversibility.~\cite{Evans93,Esposito09}
It gives an identity equation on the probability of the entropy production $\Delta S$ during time $\tau$ as
\begin{equation}
\ln \left[ \frac{P(\Delta S)}{P(-\Delta S)} \right] = \Delta S,
\end{equation}
in the asymptotic limit $\tau \rightarrow \infty$.
While this expression reproduces the ordinal Onsager relations in the linear response regime,
it provides useful information even in the far-from-equilibrium regime.

The importance of the Onsager-Casimir relation on coherent quantum transport of mesoscopic devices 
is known for long time.~\cite{Datta95etc,Buttiker86} 
Recently, Saito and Utsumi have derived a quantum version of the fluctuation theorem,
and applied it for derivation of universal relations among nonlinear transport coefficients~\cite{Saito08} by means of the full-counting statistics in the Keldysh formalism.~\cite{Keldysh64,Levitov93,Nazarov03}
In Ref.~\citen{Saito08}, however, its physical meaning on thermoelectric effects has not been addressed in detail. 

In this note, we discuss extension of the Onsager relation between the Peltier effect and 
the Seebeck effect to nonequilibrium steady states in mesoscopic devices. 
For simplicity, we consider a two-terminal setup,
though extension to a multi-terminal setup is straightforward. 
We consider the Hamiltonian $H =\sum_{r = {\rm L}, {\rm R}} H_r + H_d + H_T$, where
\begin{align}
H_r=& \sum_k \epsilon_{rk}a^\dag_{rk}a_{rk}, \\
H_d =& \sum_{i=1}^{m}\epsilon_{i}d^\dag_{i}d_{i} 
+ \sum_{\langle i,j \rangle}^{m} t_{ij} (c_i^{\dagger} c_j + {\rm h.c.})  + H_I, \\
H_T =& \sum_{rki} t_{rki} (d^\dag_{i}a_{rk} + {\rm h.c.}).
\end{align}
Each of two leads is described by $H_r$ ($r = {\rm L}, {\rm R}$),
where $a_{rk}$ is an annihilation operator of an electron with a wave vector $k$.
A mesoscopic device is modeled in a general form 
by $H_d$ consisting of an arbitrary number of local energy levels labeled by $\epsilon_i$ ($1\le i \le m$),
where $d_i$ is an annihilation operator at the $i$th site. In this model,
electron hopping $t_{ij}$ and electron-electron interaction 
(denoted by $H_I$) between arbitrary pairs of sites are assumed.~\cite{footnote1} A coupling between 
the leads and the mesoscopic device is described by $H_T$ with electron hopping $t_{rki}$.
An external magnetic field $B$ is introduced by the Peierls phase on 
the hopping elements as $t_{rki}=\left| t_{rki} \right| \exp({\rm i}\phi_{rki})$ and 
$t_{ij}=\left| t_{ij} \right| \exp ({\rm i}\phi_{ij})$ where $\phi_{rki}$ and $\phi_{ij}$ are odd functions of
the magnetic field: $\phi(-B) = -\phi(B)$. In this note, we use a unit $\hbar=k_B=e=1$.

We introduce a cumulant generating function (CGF) ${\cal F}(\chi_c,\chi_e;B)$ for the steady state,
where $\chi_c$ and $\chi_e$ are counting fields for charge and heat current.
Current operators are defined as
$I_c = \dot{N}_{\rm L} = {\rm i}[N_L,H_T]$ and $I_e = \dot{H}_{\rm L} = {\rm i}[H_L,H_T]$~\cite{footnote2},
where $N_L$ is a number of particle of the left lead.
The CGF generates cumulants by the derivatives with respect to the counting fields.
For example, the charge current and noise are generated as
\begin{align}
\langle\!\langle I_{c} \rangle\!\rangle &= \langle I_c \rangle =
\left. \frac{\partial {\cal F}}{\partial i\chi_{c}} \right|_{\chi_c = \chi_e = 0}, \\
\langle\!\langle I_{c}^2 \rangle\!\rangle &= \langle I_c^2 \rangle - \langle I_c \rangle^2 =
\left. \frac{\partial^2 {\cal F}}{(\partial i\chi_{c})^2} \right|_{\chi_c = \chi_e = 0},
\end{align}
respectively.
General relations among nonlinear transport coefficients are derived from a symmetry relation
\begin{align}
{\cal F} (\chi_{c}, \chi_{e} ;  B) = {\cal F} ( -\chi_{c}+i{\cal A}_{c}, -\chi_{e}+i{\cal A}_{e};  -B),
\label{symmetry}
\end{align}
which is a consequence of  the microscopic reversibility~\cite{Saito08}.
Here, ${\cal A}_{c} = \beta_L\mu_L-\beta_R\mu_R$ and ${\cal A}_{e} = - (\beta_L - \beta_R)$ are affinities. 
All the cumulants are expanded with respect to ${\cal A}_c$ and ${\cal A}_e$ as
\begin{align}
\langle \! \langle I_c^{k_1} I_e^{k_2} \rangle \! \rangle 
&= \sum_{l_1=0}^{\infty} \sum_{l_2=0}^{\infty} \frac{L_{l_1, l_2}^{k_1,k_2}}{l_1! l_2!} {\cal A}_c^{l_1} {\cal A}_e^{l_2}, \\
L^{k_1,k_2}_{l_1,l_2}\left(B\right) 
&= \left. \frac{\partial^{l_1+l_2}\langle\!\langle I^{k_1}_{c}I^{k_2}_{e} \rangle\!\rangle}
{ \partial {\cal A}^{l_1}_{c} \partial {\cal A}^{l_2}_{e} } \right |_{{\cal A}_{c}={\cal A}_{e}=0}.
\end{align}
By symmetrizing and anti-symmetrizing the transport coefficients with respect to $B$ as 
\begin{align}
L^{k_1,k_2}_{l_1,l_2,\pm}\left(B\right) 
&= L^{k_1,k_2}_{l_1,l_2}\left(B\right) \pm L^{k_1,k_2}_{l_1,l_2}\left(-B\right),
\end{align}
general relations are derived from Eq.~(\ref{symmetry}) as~\cite{Saito08}
\begin{align}
\nonumber
L^{k_1,k_2}_{l_1,l_2,\pm}
&= \pm \sum_{n_1=0}^{l_1} \sum_{n_2=0}^{l_2} \binom{l_1}{n_1} \binom{l_2}{n_2} \\
&\times \left(-1\right)^{n_1+n_2+k_1+k_2} L^{k_1+n_1,k_2+n_2}_{l_1-n_1,l_2-n_2,\pm}.
\label{RelOfCo}
\end{align}
We can show immediately that Eq.~(\ref{RelOfCo}) includes the Onsager relation in the linear response region
as follows. 
First, $L_{l_1, l_2}^{k_1, k_2}(B)$ are related to ordinary observables
by changing affinities to the external-bias variables,
${\cal A}_c$ and ${\cal A}_e$ to 
the bias voltage $V = \mu_L - \mu_R$ and the temperature difference $\Delta T = T_L - T_R$. (For example, 
the linear conductance of charge current, the charge current noise at equilibrium and the Seebeck coefficient are written as 
$G = L_{1,0}^{1,0}/T$, $S_{I, I} = L_{0,0}^{2,0}$ and $S= L_{0,1}^{1,0}/ TL_{1,0}^{1,0}$,
 respectively.) 
We note that the linear response of charge and heat currents are
expressed by $I_c = L_{1,0}^{1,0}(B) (V/T) + L_{0,1}^{1,0}(B) (\Delta T/T^2)$ and 
$I_e = L_{1,0}^{0,1}(B) (V/T) + L_{0,1}^{0,1}(B) (\Delta T/T^2)$.~\cite{Callen85}
Then, specific equations between coefficients satisfying 
$N \equiv k_1+k_2+l_1+l_2 = 2$ lead to the Onsager relation $L_{0,1}^{1,0}(B) = L_{1,0}^{0,1}(-B)$.

Next, we prove that the magnitude of the nonlinear Peltier effect can be determined only by
information of charge current measurements without heat current measurement;
we show that all the coefficients for heat current appearing in the expansion
\begin{align}
I_e = \sum_{l_1=0}^{\infty} \frac{L_{l_1,0}^{0,1}}{l_1!} {\cal A}_c^{l_1} = L^{0,1}_{1,0}{\cal A}_c+L^{0,1}_{2,0}{\cal A}_c^2/2+ 
{\cal O}({\cal A}_c^3) ,
\label{heati}
\end{align}
under the isothermal condition (${\cal A}_e = 0$) can always be rewritten by the transport coefficients
of the higher-order charge current cumulant $L_{l_1,1}^{k_1,0}$. By substituting $k_2 = 0$ and $l_2 = 1$
into the general equation (\ref{RelOfCo}), we obtain
\begin{align}
L_{N-i , 1, \pm}^{i , 0} = \pm  \sum_{j=0}^N M_{ij} \left( L_{N-j, 1,\pm}^{j , 0} - L_{N-j , 0,\pm}^{j , 1} \right),
\end{align}
where $M$ is a matrix whose matrix elements are given as 
\begin{align}
M_{ij} = \left\{
\begin{array}{cc}
	\displaystyle
	\binom{N \!-\! i}{j \!-\! i} (-1)^j & (0 \le i \le j \le N) \\
	\displaystyle 0 & (0 \le j < i \le N) 
\end{array}\right. .
\end{align}
By utilizing $\sum_j M_{ij} M_{jk} \! = \! \delta_{ik}$, we obtain
\begin{align}
L_{N ,0,\pm}^{0 , 1} = \mp \sum_{k=1}^N M_{0 k} L_{N-k , 1,\pm}^{k , 0}.
\label{eq1}
\end{align} 
We notice that Eq.~(\ref{eq1}) relates the nonlinear response of the heat current under ${\cal A}_e = 0$
to that of the charge current, current noise, and higher cumulants up to the linear response in ${\cal A}_e$.
This means that the magnitude of nonlinear Peltier effect can be evaluated without direct measurement of the heat current.
This result can be regarded as a nonequilibrium extension of the Kelvin-Onsager relation.

Finally, we discuss the lowest-order nonlinear correction of the Peltier coefficient defined by
\begin{align}
\Pi({\cal A}_c) = \frac{I_e}{I_c} = \Pi^{(0)} + \Pi^{(1)} I_c + {\cal O}(I_c^2) . 
\end{align}
under the isothermal condition (${\cal A}_e = 0$).
The linear-response term satisfies the Kelvin-Onsager relation $\Pi^{(0)}=TS^{(0)}(-B)$ as already mentioned,
where $S^{(0)}(B)$ is the ordinal Seebeck coefficient defined in the linear-response regime.
The first-order correction $\Pi^{(1)}$ is formally expressed by using the definition of $L_{l_1,l_2}^{k_1,k_2}$ as 
\begin{align}
\Pi^{(1)} = \frac{L^{0,1}_{1,0}(B)}{2 L^{1,0}_{1,0}(B)^2}
\left( \frac{L^{0,1}_{2,0}(B)}{L^{0,1}_{1,0}(B)} - \frac{L^{1,0}_{2,0}(B)}{L^{1,0}_{1,0}(B)} \right). 
\nonumber 
\end{align}
By utilizing the general relation in Eq. (\ref{RelOfCo}) for $N=2$, the correction is rewritten as
\begin{align}
\frac{\Pi^{(1)}}{T}  =&  - \frac{{S}(-B)}{2}
\biggl( \frac{L^{1,0}_{2,0}(B)}{L^{1,0}_{1,0}(B)^2} 
\nonumber \\
+& \frac{{L}^{2,0}_{0,1}(-B)-2 {L}^{1,0}_{1,1}(-B)}{L^{1,0}_{1,0}(B) {L}^{1,0}_{0,1}(-B)} \biggl) \, .
\label{nonlinearko}
\end{align}
Thus, though the expression is a little complicated, one can find that 
the nonlinear correction of the Peltier coefficient can be calculated only by coefficients $L_{l_1, l_2}^{k_1, 0}$,
which needs no heat-current measurements.
In a similar way, higher-order correction of the Peltier coefficient $($e.g. $\Pi^{(2)})$
can be expressed by higher-order cumulants of the charge current (e.g. skewness).

We proved that the magnitude of the nonlinear Peltier effect under the isothermal condition can be evaluated from
transport coefficients for charge current cumulants (current, current noise, and higher cumulants) up to the linear
response to a thermal bias.
Inversely, if precise measurement of both charge and heat currents is possible, thermoelectric effects in the nonlinear regime
can be used for an experimental proof of the fluctuation theorem in quantum systems, as performed
in a recent experiment by means of simultaneous measurement of conductance and shot noise.~\cite{Nakamura09}
By combining recent progress in measurement techniques of thermoelectric effects,~\cite{Scheibner07} 
the present result will provide an important key to understand nonequilibrium thermoelectric effects in mesoscpic systems.

We would like to thank Keiji Saito
for drawing our attention to the FT and the  thermal transport. 
Y. U. acknowledges financial support by Strategic International Cooperative Program JST.
E. I. and T. K. acknowledge financial support by JSPS and MAE
under the Japan-France Integrated Action Program (SAKURA). This work was supported by
Grant-in-Aid for Young Scientists (B) (21740220).


\begin{thebibliography}{99}
\bibitem{footnote1}
We note that the model considered here can cover a large number of mesoscopic systems,
e.g., including Coulomb interaction $H_I = \sum V_{ij} n_i n_j$ ($n_i = d^{\dagger}_i d_i$)
and spin degrees of freedom (not explicitly written in this note).
\bibitem{footnote2}
Because the leads consists of noninteracting electron systems, the energy current
coincides with the heat current in the present definition.
\end{thebibliography}
\end{document}